\documentclass[12pt]{iopart}

\usepackage{iopams}
\usepackage{graphicx}

\begin{document}


\title{Octave-Spanning Phase Control for Single-Cycle Bi-Photons}

\author{Yaakov Shaked, Shai Yefet, Tzahi Geller and Avi Pe'er$^{*}$}

\address{
Department of physics and BINA Center for nano-technology, Bar-Ilan university, Ramat-Gan 52900, Israel

$^*$e-mail: avi.peer@biu.ac.il
}

\begin{abstract}
The quantum correlation of octave-spanning time-energy entangled bi-photons can be as short as a single optical cycle. Many experiments designed to explore and exploit this correlation require a uniform spectral phase (transform-limited) with very low loss. So far, transform-limited single-cycle bi-photons were not demonstrated, primarily due to lack of precision control of the spectral phase. Here, we demonstrate precise correction of the spectral-phase of near-octave bi-photons to less than ($<\pi/20$) (residual phase) over nearly a full octave in frequency ($\approx1330-2600$ nm). Using a prism-pair with an effectively-negative separation, we obtain tuned, very low-loss compensation of both the 2nd and 4th dispersion orders. We verify the bi-photons spectral phase directly, using a non-classical bi-photon interference effect.
\end{abstract}



\noindent The ultra-broad bandwidth of time-energy entangled photons that are generated from a narrowband pump (bi-photons) enfolds a unique resource for quantum information. Due to the ultra-short correlation in time, yet continuous generation, an ultra-high flux of single bi-photons can be generated, which is specifically attractive for high-speed quantum communication and quantum cryptography \cite{TeleportaionGisin,1367-2630-16-3-033017}. Single-cycle bi-photons, whose bandwidth covers an octave in frequency and temporal correlation is comparable to the optical cycle, are the extreme realization of time-energy entanglement, and a long-sought goal in experimental quantum optics [ref. Harris and others].
In the pursuit of single-cycle bi-photons, two main issues need to be addressed - bi-photon generation with sufficient bandwidth, and precise spectral-phase control to overcome the dispersive temporal-broadening of the correlation and compress it to a single-cycle duration. In addition, low photon loss is critical, as the non-classical correlation is inherently sensitive to it. Generation of octave spanning bi-photons requires ultra-broad phase matching of the nonlinear down-conversion interaction, and has been widely discussed  \cite{SingleCycleTheoryHarris,1367-2630-16-5-053012}. In \cite{ChirpCompressBiPhotonsHarris,ExperimentChirpCompressBiPhotonHarris,SingleCycleBiPhotonsKolobov} for example, the broad bandwidth is achieved by tailoring the quasi-phase matching of the process using a modulated periodic polling (chirped polling). We have recently demonstrated a simple method for generating ultra-broad bi-photons by type-I down conversion near the zero-dispersion frequency of the non-linear crystal (cite our paper), with a near octave-bandwidth (over 100THz between 120-220 THz).
Precise control of the spectral phase has been far less addressed. When the spectral phase of the bi-photons is modulated by group velocity dispersion in matter, or any other means, the bi-photons are no longer transform-limited and the major bi-photon measurement methods, Hong-Ou-Mandel (HOM) interference \cite{PhysRevLett.59.2044}, and sum frequency generation (SFG) \cite{PhysRevLett.98.063602,PhysRevLett.94.043602,PhysRevLett.75.3426,1367-2630-16-3-033017}, which are both broadband interference effects that are highly sensitive to phase modulation, are no longer applicable without compensating the dispersion. Compensation of the spectral phase over such an ultra-broad bandwidth is not trivial. The spectral phase shaping mechanism needs to be very low-loss in order to maintain the quantum bi-photon state, and to provide compensation of at least two orders of dispersion (2nd and 4th) to fully exploit the bi-photon bandwidth. In addition, as in many experiments, when the bi-photons are generated in the IR range, where the dispersion of most optical materials becomes negative, standard techniques for dispersion control in the visible and near-IR range are no longer applicable, such as a prism-pair \cite{prismpairapprox} or grating-pair \cite{gratingpair}, which inherently rely on the negative dispersion of free space between dispersive elements to compensate for the generally positive dispersion of optical materials. When moving deeper into the IR range, this concept must be expanded to allow compensation of also negative material dispersion.

Here, we describe and demonstrate compensation of the bi-photon spectral phase to $<\!\pi/20$ accuracy across a bandwidth of $>\!110$ THz (nearly an octave). Introducing an effectively \emph{negative} separation we are able to utilize a Brewster-cut prism-pair \cite{prismpairapprox},to compensate simultaneously the two dispersion orders: 2nd (group delay dispersion, GDD) and 4th (fourth order dispersion, FOD). Note, that the phase of a single photon in the pair is undefined and only the phase-sum of both photons is of interest (correlated to the pump phase). Thus, compensation is needed only for even (symmetric) orders of the dispersion \cite{1303721,Christov:94,Cojocaru:03}, while odd (anti-symmetric) orders have no influence, as they leave the phase-sum unaffected. Nonetheless, the compensation used here can be relevant for odd orders of dispersion also, where necessary.

Our broadband bi-photon source relies on SPDC, where the bandwidth of the bi-photons is limited only by phase matching, indicating that ultra-broadband SPDC can be obtained if the pump frequency coincides with the zero-dispersion of the nonlinear crystal. We use a periodically-poled KTP (PPKTP) crystal, pumped by a narrow-band laser at $880$nm. The generated SPDC is symmetrical in frequency around the degenerate wavelength, $\lambda_{0}=1760$nm (twice the pump), which is nearly the zero dispersion at $\lambda_{0}=1790$nm. With this method, we generate bi-photons spanning from $\approx115$THz to $\approx225$THz around the degenerate frequency of $\nu_{0}=170$THz (nearly an octave). The phase variation of the SPDC due to the residual phase mismatch in the crystal is the most fundamental dispersion that needs compensation, in addition to dispersion from other optical elements in the setup.
\begin{figure}
\centerline{\includegraphics[clip, trim=0cm 0cm 0cm 0cm, height=3cm, width=8.25cm]{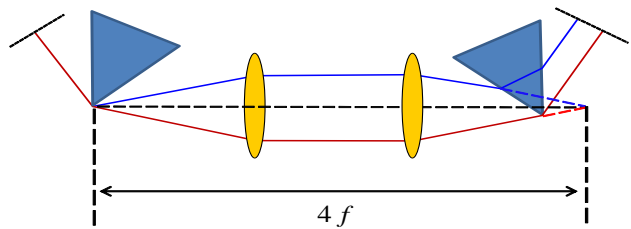}}
\caption{\label{BPPsystem}  A prism-pair with negative separation. The telescope images the vertex of the 1st prism forward (distance $4f$). Placing the second prism before the image results in an effective \emph{negative} separation $R$ between the prisms.}
\end{figure}

A common simple technique for dispersion compensation uses a pair of Brewster-cut prisms \cite{Sherriff:98}. By varying the separation $R$ between the prisms tips and insertion $H$ of the prisms into the beam path, the geometrical and material dispersion can be dynamically controlled to tune the overall dispersion. The prism-pair is preferable to other techniques of dispersion control due to its ultra-low loss and high degree of real-time tunability. A grating-pair, for example, can compensate only for one order of dispersion (GDD) and is generally lossy due to the non-ideal diffraction efficiency. Higher-order compensation with gratings requires introduction of pulse shaper \cite{Weiner20113669}, which is costly, complicated and incurs additional loss. Note that for bi-photons, loss is specifically deleterious, since loosing only one photon already implies loss of the pair.

In most ultrafast applications, the prisms material GDD is positive, whereas the geometrical dispersion created by the separation between the prisms \emph{always} introduces negative GDD. This technique is therefore most suitable for dispersion compensation in the visible and near IR spectrum, where most optical materials produce positive GDD. In our case, however, the broad bi-photons spectrum is in the short-wavelength infra-red (SWIR) range and beyond, where most optical materials produce \emph{negative} GDD. Thus, the separation of the prisms cannot compensate for material dispersion, considerably limiting the choice of materials that can match the experimental needs and posing a major hurdle for the prism-pair to produce an overall compensation.

In some cases, a delicate balancing of the two dispersion knobs - separation $R$ and insertion $H$ of the prisms, enables compensation of \emph{two} orders of dispersion, such as GDD and TOD. However, such a simultaneous compensation of two orders is not guaranteed, as it may require a configuration with either negative separation between the prisms or a negative penetration of the prism into the beam. Indeed, we found that for compensating both GDD and FOD of our bi-photons, the solution requires a \emph{negative} separation between the prisms. Note that negative separation can be physical if we introduce an imaging system between the two prisms \cite{Weiner:88,gratingtelescope}, which images the first prism tip beyond the location of the second prism, as illustrated in Fig.\ref{BPPsystem}.

Obviously, the introduction of a negative separation inverts the sign of the geometrical dispersion and enables the prism-pair to produce a total positive GDD, even for prism material with negative dispersion. Furthermore, a negative distance provides a new knob for dispersion management $-$ the ability to tune not only the magnitude but also the sign of any order of the geometrical dispersion across a wide range, allowing optimization of the optical phase by exploiting the interplay between different orders. In our experiment, this allows compensation of both GDD and FOD, which is impossible with a standard, positively separated prism-pair or with a grating-pair \cite{gratingpair,Martinez:84}. Our analysis of the frequency dependent optical path in the prism-pair relies on a generalization of the original method presented by Fork \cite{prismpairapprox} and is detailed in \cite{1411.0232}.

\begin{figure}
\centerline{\includegraphics[height=7cm, width=8cm]{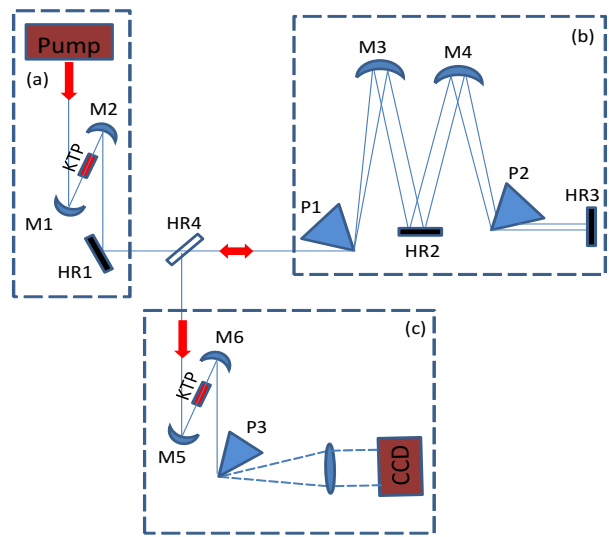}}
\caption{\label{apparatus} Experimental setup: collinear SPDC is generated in the first KTP crystal (marked (a)) and double passes through (b) a prism-pair setup (P1 and P2, Sapphire) with an intermediate 4f telescope consisting of two spherical mirrors (M3 and M4, $f=500$mm). The compensated SPDC spectrum and the $880$nm pump exit at a slightly lower hight than the input, and are directed by a lower mirror (HR4) into a second identical KTP crystal (c) where the nonclassical interference occurs. The focusing spherical mirrors around the KTP crystals (M1, M2, M5 and M6) have a focal length of $f=75$mm. All the mirrors in the setup (spherical and plane) are metallic coated, either silver or gold. The spectrometer consists of a third sapphire prism (P3), a focusing lens ($f=125$mm) and a cooled CCD camera (SWIR range).}
\end{figure}

We measure experimentally the bi-photon spectral phase with the non-classical interference effect presented in \cite{1367-2630-16-5-053012} which is capable of measuring the bi-photon spectral phase and amplitude even at the presence of a spectral modulation. The SPDC generated in one non-linear crystal propagates together with the pump laser into a second identical crystal, where SPDC can be either enhanced or diminished. Quantum mechanically, the two possibilities to generate bi-photons (either in 1st crystal or the 2nd) interfere according to the relative phase between the pump and the SPDC acquired between the crystals. Two types of relative-phase are possible: 1. the pump itself acquires a phase relative to the entire bi-photon spectrum, and 2. a spectrally varying phase over the SPDC spectrum. The first leads to a intensity variation of the entire spectrum together, whereas the second leads to the appearance of interference fringes across the spectrum. By analyzing the spectral interferogram, we can reconstruct the spectral phase. When the dispersion is fully compensated, the spectral phase becomes flat and only uniform variation of the entire spectrum will be observed without any spectral fluctuation, when the pump phase is varied.
\begin{figure}
\centerline{\includegraphics[height=6cm, width=8cm]{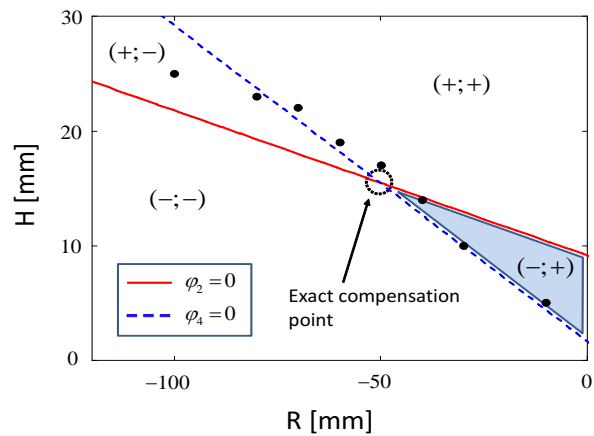}}
\caption{\label{contourmap} Zero-level contour lines of the 2nd and 4th order terms in ($R,H$) space for $\lambda_{0}=1760$nm. The $\pm$ signs in parenthesis represent the sign of the 2nd and 4th order terms, correspondingly, at the relevant sector in the contour map. The sign of the 6th dispersion order in the vicinity of the exact compensation point is negative, hence a compromise in the values of the 2nd and 4th order dispersion terms within the (-,+) sector will allow broader compensation. The black filled circles represent the points where maximum compensated bandwidth was achieve for a given value of $R$.}
\end{figure}

The experimental configuration is illustrated in Fig.\ref{apparatus} and consists of three main parts: First, ultra broadband bi-photons are generated via collinear SPDC in a Brewster cut $12$mm-long, PPKTP crystal  pumped by a single-frequency diode laser at 880nm (of power $\approx0.5$W). The SPDC produces $\approx10^{12}$ bi-photons per second with a bandwidth of $\approx100$THz. The generation probability of bi-photons $\approx1/90$  photon-pairs per Hertz per second. Next, the generated bi-photons propagate through a folded prism-pair system (double-pass), where a pair of sapphire prisms are separated by a reflective telescope constructed from two gold spherical mirrors. By double-passing the prism-pair system, we guarantee exact re-packaging of the spectrum. In our experiment, the negative distance $R$ between the tips is a few centimeters while the prisms insertion $H$ is several mm. The shaped spectrum enters a 2nd identical crystal (along with the pump) where bi-photon interference can occur in the form of enhanced/diminished bi-photon generation (constructive/destructive interference). The generated bi-photon spectrum and spectral fringes are measured with a home-built spectrometer composed of a sapphire prism coupled to a cooled CCD camera. A symmetrical spectral interference pattern is observed with contrast of $15-20\%$, which are then used to extract the spectral phase acquired by the SPDC light before the second crystal.
\begin{figure}
\centerline{\includegraphics[height=6cm, width=8cm]{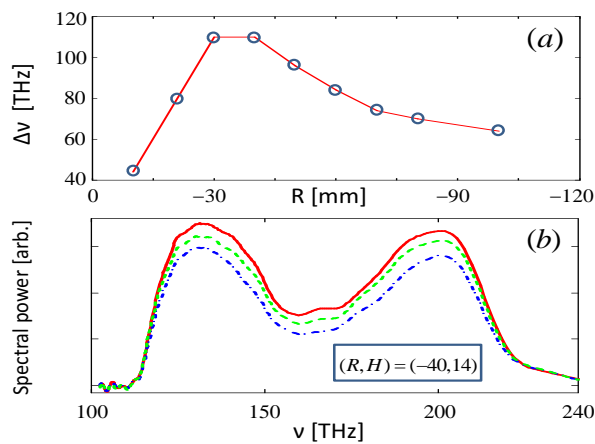}}
\caption{\label{bandwidth} (a) Measured compensated bandwidth $\Delta\nu$ as a function of the negative prism separation $R$. (b) Spectral power vs. frequency at the optimal point of compensation. The curves represent destructive, intermediate and constructive interference, respectively (with visibility of $\approx15\%$). The interference is evidently uniform across the entire spectrum.}
\end{figure}

We now wish to find the optimal dispersion compensation for the bi-photons among the different possible ($R,H$) configurations of the prisms (separation and penetration). Optimal compensation should be obtained near the calculated vanishing point of both the GDD and FOD (see Fig.\ref{contourmap}). We therefore varied $R$ and $H$ according to the following protocol: we scanned the separation $R$ from a large negative separation towards $R=0$. For each separation $R$, we tuned the prism insertion $H$ to achieve the broadest possible compensation at this $R$ and estimated the compensated bandwidth by measuring the bandwidth over which the spectral interference variation was uniform (see details later on). Figure \ref{bandwidth}(a) depicts the measured compensated bandwidth as a function of the negative separation $R$, showing optimal compensation at $R\approx-40$mm with the prisms insertion $H\approx14$mm. Initially, we aimed for a maximal phase fluctuation of $\Delta\varphi<\pi/10$ across the entire spectrum, but analysis shows that the compensated fluctuation was better ($\Delta\varphi\approx\pi/20$) over the entire $>110$THz bandwidth. Figure \ref{bandwidth}(b) shows the measured spectrum at the optimal point, demonstrating the uniform variation of the entire spectrum between destructive and constructive interference.

In conclusion, we demonstrated dispersion compensation of ultra broadband bi-photons, using a prism-pair with an effectively negative separation. The compensated bandwidth was measured by a nonlinear pair-wise interference. The low-loss ultra-broad phase compensation will enable utilization of broadband quantum measurement methods, such as HOM and SFG, opening an avenue to quantum optics applications with ultra-broadband, high-flux bi-photons.


Last, we provide details on the retrieval of the residual phase fluctuation in the experiment. The spectral interferogram intensity can be expressed as $I(\omega)=I_{0}(\omega)[1+V(\omega)\cos(\varphi_{0}+\varphi(\omega))]$\, where $I_{0}(\omega)$ is the average spectrum without interference, $V(\omega)$ is the fringe visibility at frequency $\omega$ (of order $15-20\%$ in our experiment), $\varphi_{0}$ is the overall phase of the pump and $\varphi(\omega)$ is the spectral phase variation of interest.

In order to retrieve the spectral phase variation, we normalize the interferogam according to $I_{norm}=I(\omega)/I_{0}(\omega)-1=V(\omega)\cos(\varphi_{0}+\varphi(\omega))$ and assume a constant fringe contrast $V(\omega)=V_{0}$. The phase variation is most pronounced when the phase of the pump is $\varphi_{0}\approx\pi/2$, where $\cos(\varphi_{0}+\varphi(\omega))\approx\varphi(\omega)$ (assuming a small phase variation $\varphi(\omega)$), indicating that the spectral phase is $\varphi(\omega)=I_{norm}/V_{0}$.

\begin{figure}
\centerline{\includegraphics[height=6cm, width=8cm]{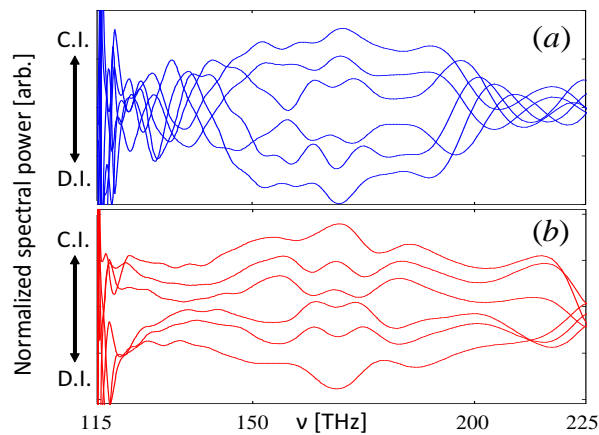}}
\caption{\label{interference} Normalized spectral intensity, illustrating two examples of the nonclassical pairwise interference pattern for two different prism-pair configurations: a narrow (a) and broad (b) compensated bandwidth. The corresponding values of the prism-pair parameters $R$ and $H$ (in mm) for each plot are $(R,H)=(-10,5)$ for (a) and $(R,H)=(-40,14)$ for (b).}
\end{figure}

Figure \ref{interference} shows representative sets of normalized interferograms for non-optimal compensation (Fig.\ref{interference}(a)) and at the optimal point (Fig.\ref{interference}(b)), where the entire spectrum apparently varies uniformly. In order to demonstrate full compensation, we have recorded a large set of normalized interferograms at various $\varphi_{0}$ and analyzed specifically those of $\varphi_{0}\approx\pi/2$. For Fig.\ref{interference}(b), analysis indicates a residual phase variation of $<\pi/20$.

This research was supported by the Israeli Science Foundation (grant 44/14).

\bibliographystyle{plain}

\end{document}